\documentclass[12pt]{article}
\usepackage{amssymb,amsmath,epsfig}

\begin{document}

\title{\bf Geodesic Study of Regular Hayward Black Hole}

\author{G. Abbas \thanks{ghulamabbas@ciitsahiwal.edu.pk} and U.
Sabiullah
\thanks{umairmughal70@gmail.com}\\
Department of Mathematics, COMSATS Institute,\\
of Information Technology, Sahiwal-57000, Pakistan.}

\date{}
\maketitle
\begin{abstract}
This paper is devoted to study the geodesic structure of regular
Hayward black hole. The timelike and null geodesic have been studied
explicitly for radial and non-radial motion. For timelike and null
geodesic in radial motion there exists analytical solution, while
for non-radial motion the effective potential has been plotted,
which investigates the position and turning points of the particle.
It has been found that  massive particle moving along timelike
geodesics path are dragged towards the BH and continues move around
BH in particular orbits.
\end{abstract}

\

\section{Introduction}

Since yet, we do not have perfectly reliable candidate to test the
phenomenological aspects of quantum theory of gravity. Spacetime
singularity is one of the avoidable issue in quantum theory of
gravity, so research on properties of regular black hole solutions
is much important in quantum theory of gravity. The regular black
hole solutions have event horizons but there is the absence of such
region where closed timelike curves may exist, i.e., such solutions
are singularity free. Of course, these are non-vacuum solutions of
Einstein field equations and source of gravity is taken as some form
of exotic field, non-linear electrodynamics or some modified form of
gravity.

The regular BH solution was formulated by Bardeen which is commonly
known as Bardeen BH (Bardeen 1968, Borde 1994 and Borde 1997). Later
on, Ayon-Beato and Garcia (Ayon-Beato and Garcia 1998) carried out
the coupling of Einstein field equation and Maxwell field equations
and derived the charged version of Bardeen. For this version of
regular BH, they take the non-linear electric field as a source of
charge for the solution of Maxwell field equations. The physical
nature of charged regular BH was studied by the Bronikov (Bronnikov,
2000, Bronnikov, 2001), he showed analytically that charged regular
BHs are not correct solutions of the field equations. The main cause
of such unsatisfactory interpretation was the addition of
electromagnetic Lagrangian quite differently in the various
direction of the space under consideration. On the other hand a
correct solution of the field equations was given by Hayward
(Hayward, 2006), which is free of charge term and its physical
aspects are quite similar to Bardeen BH. The limits of the Hayward
BH are quite correct as in the limit $r\rightarrow\infty$, it
corresponds to Schwarzschild BH and for $r\rightarrow0$, it
corresponds to de-Sitter BH. The particular choice of solution
parameters imply that it may have two horizons, single horizon and
no horizon. A lot of work (Sharif and Abbas 2013a, 3013b, 2013c) has
been done on the nature and structure of regular BH.

One of the most challenging tasks in theoretical physics is to
combine quantum theory and general relativity, which can be done by
clarifying the structure of singularities occurring in general
theory of relativity. In this regard the geodesic study near the
gravitational field of compact objects is much important. This study
was initiated by Chandrasekhar (Chandrasekhar 1983), he studied the
geodesic of Schwardzschild, Reissner-Nordstr$\ddot{o}$m and Kerr
BHs. Since then there has been growing interest to study the
geodesic around BHs. Many researches have studied the geodesics of
various BH like geodesic study of the schwarzschild BH in Rainbow
gravity was studied by Leiva et al.(2009). Guha and Bhattacharya
(2012) the geodesic motion around five dimensional charged anti-de
Sitter BH. Fernando et al.(2003) have studied the null geodesic of
Schwarzschild BH in the presence of quintessence.

Recently, Kalam et al.(2014) have explored the geodesic of charged
BH in non-linear electrodynamics. Curz et al.(2013) have studied the
geodesic structure of topological Lifshitz black hole in 1+2
dimension. Mosaffa (2011) have explored the geodesic structure in
Horava-Lifshitz BH. Setare and Mansouri (2003) have investigated the
null geodesics in Horava-Lifshitz gravity. Several authors have
discussed the geodesics of BHs with cosmological constant. In this
paper, we have studied the geodesic of a regular Hayward BH, which
has properties of Schwarzschild as well as de-Sitter BH.

\section{Geodesics of Regular Hayward Black Hole}

The static spherically symmetric space-time is described by the
Hayward metric (Hayward 2006) which is given by
\begin{equation}\label{1}
{ds}^2=-f(r)dt^2+\frac{1}{f(r)}
dr^2+r^2(d\theta^2+\sin^2{\theta}d\phi^2),
\end{equation}
where $f(r)=1-\frac{2mr^2}{r^3+2l^2m}$ with $m$ corresponding  to
mass of the black hole and $l$ is a convenient encoding of the
central energy density $3/8\pi l^{2}$, assumed positive. The lapse
function $f(r)$ for $\lim r\rightarrow\infty$ reduces to
$1-\frac{2m}{r}+O(\frac{1}{r^4})$, while at $\lim r\rightarrow0$ it
reduces to $1-\frac{r^2}{l^2}+O(r^4)$. From the asymptotic behavior
of the metric function, it is clear that Hayward BH becomes
Schwarzschild BH at large value of $r$ and for small value of $r$,
it is de-Sitter BH. The Hayward BH solution is non-singular
(regular) as all the scalars curvature are finite and regular at
center of the metric where $r\rightarrow0$ , which can be verified
by scalars of this metric
\begin{eqnarray}\label{1}
R&=&\frac{24l^2m^2(r^3-4l^2m)}{(r^3+2l^2m)^3},\quad
{\lim}_{r\rightarrow0
}R=-\frac{12}{l^2},\\\
R_{\mu\nu}R^{\mu\nu}&=&\frac{288m^4l^4(5r^6-6r^3l^2m+8l^4m^2)}{(r^3+2ml^2)^6},\quad
 {\lim}_{r\rightarrow0
}R_{\mu\nu}R^{\mu\nu}=\frac{36}{l^4},\\\nonumber
R_{\mu\nu\lambda\eta}R^{\mu\nu\lambda\eta}&=&\frac{48m^2}
{(r^3+2ml^2)^6}\left(r^{12}-8r^9ml^2+72m^2r^6l^4-16r^3m^3l^4+32m^4l^8\right)\\\
{\lim}_{r\rightarrow0
}R_{\mu\nu\lambda\eta}R^{\mu\nu\lambda\eta}&=&\frac{24}{l^4}.
\end{eqnarray}

This metric function $f(r)$ reduces to the Schwarzschild BH for
$l=0$ and is flat for $m=0$. By the analysis of $f(r)$ for zeros, we
see that a critical mass $m_{*}=(3\sqrt{3}/4)l$ and
$r_{*}=\sqrt{3}l$ such that $r>0$,  $f(r)$ has no zero if $m<m_{*}$
and if $m=m_{*}$ there is one zero at $r=r_{*}$ and if $m>m_{*}$
there are two zeros at $r=r_{\pm}$, the event and inner horizons.
This is shown in figure \textbf{1}.
\begin{figure}
\begin{center}
  \includegraphics[width=11cm]{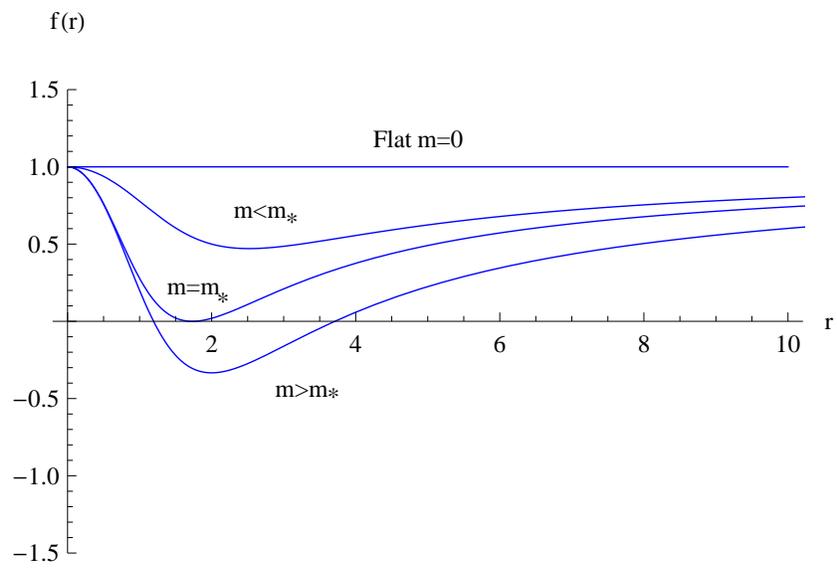}\\
  \caption{Behavior of $f(r)$ for fixed value of parameter $l$ and different values of
  $m$}.\label{1}
  \end{center}
\end{figure}

Now, the Lagrangian for metric (\ref{1}) is
\begin{eqnarray}\label{2}
{\mathcal{L}}=-\frac{1}{2}\left(-f(r)\dot{t}^2+\frac{1}{f(r)}\dot{r}^2+r^2\dot{\theta}^2+r^2\sin^2{\theta}\dot{\phi}^2\right),
\end{eqnarray}
where dot indicates the differentiation with respect to affine
parameter $\tau$.

The Euler-Lagrange equation is
\begin{equation}\label{ll3}
\frac{d}{d\tau}\left(\frac{\partial {\mathcal{L}}}{\partial
\dot{x}^\mu}\right)-\frac{\partial {\mathcal{L}}}{\partial x^\mu}=0,
\end{equation}
using Eq.(\ref{2}) in Eq.(\ref{ll3}), we get
\begin{equation}\label{ll4}
\dot{t}=\frac{E}{f(r)},
\end{equation}
\begin{equation}\label{ll5}
r^2\sin^2{\theta}\dot{\phi}=J,
\end{equation}
where $E$ and $J$ are constant of motion which correspond to the
Killing vectors $\partial_t$ and $\partial_\phi$, respectively.
Further, we take $\theta=\frac{\pi}{2}$ and $\dot{\theta}=0$ as
initial conditions. Hence Eqs.(\ref{ll4}) and (\ref{ll5}) yield
\begin{equation}\label{3a}
\dot{t}=\frac{E}{f(r)},
\end{equation}
\begin{equation}\label{3b}
r^2\dot{\phi}=J.
\end{equation}
Using Eqs.(\ref{3a}) and (\ref{3b}) in Eq.(\ref{2}),
\begin{eqnarray}\label{3}
\frac{1}{f(r)}\left(\frac{dr}{d\tau}\right)^2=\frac{E^2}{f(r)}-\frac{J^2}{r^2}-L,
\end{eqnarray}
where $L=2{\mathcal{L}}$ and $L$ has values 0 and 1.
 Equation (\ref{3}) becomes for radial motion
\begin{equation}\label{4}
\frac{1}{f(r)}\left(\frac{dr}{d\tau}\right)^2=\frac{E^2}{f(r)}-L.
\end{equation}
Here,
\begin{equation}\label{5}
\frac{dr}{dt}=\frac{d\tau}{dt}\frac{dr}{d\tau}
\end{equation}
\begin{equation}\label{6}
\left(\frac{dr}{dt}\right)^2=\left.
f^2(r)\left[1-\frac{L}{E^2}f(r)\right]\right..
\end{equation}
This is the master equation for the radial geodesic motion. In
following, we shall apply this equation explicitly for photon-like
particles $L=0$ and massive particles $L=1$.

\subsection{Photon-like Particle Motion ($L$=0)}
 Equation (\ref{6}) gives

\begin{equation}\label{7}
\left(\frac{dr}{dt}\right)^2={f(r)}^2,
\end{equation}
after putting the value of $f(r)$, we get

\begin{equation}\label{9}
 \pm t=\int\frac{dr}{{1-\frac{2mr^2}{r^3+2l^2m}}}.
\end{equation}
Integration of above equation leads to
\begin{eqnarray}\label{10}
 \pm t&=&
 \frac{1}{9}\left[\frac{(42)^{2/3} m^{5/3}\log(2l^{2/3} \sqrt[3]{m} +2^{2/3}
 r)}{l^{2/3}}-\frac{(42)^{2/3}\sqrt{3}m^{5/3}\tan^{-1}\left(
 \frac{\frac{2^{2/3}r}{l^{2/3}\sqrt[3]{m}}-1}{\sqrt{3}}\right)}{l^{2/3}}\right.\nonumber\\
 &+&\left.9r+\frac{12m^{2}r^{2}}{2l^{2}m+r^{3}}+6m\log{(2l^{2}m+r^{3})}\right.\nonumber\\
&-&
\left.{\frac{(22)^{2/3}m^{5/3}\log{\left(2l^{4/3}m^{2/3}-2^{2/3}l^{2/3}\sqrt[3]{m}r+\sqrt[3]{2}r^{2}\right)}}{l^{2/3}}}\right].\nonumber\\
\end{eqnarray}
The relation between the time and distance is shown in left graph of
figure \textbf{2}.
\begin{figure}
\epsfig{file=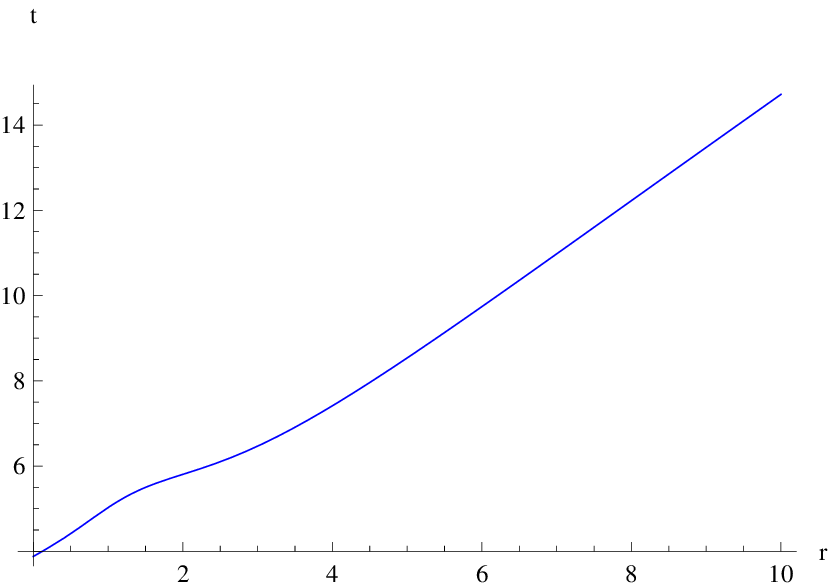, width=0.5\linewidth}\epsfig{file=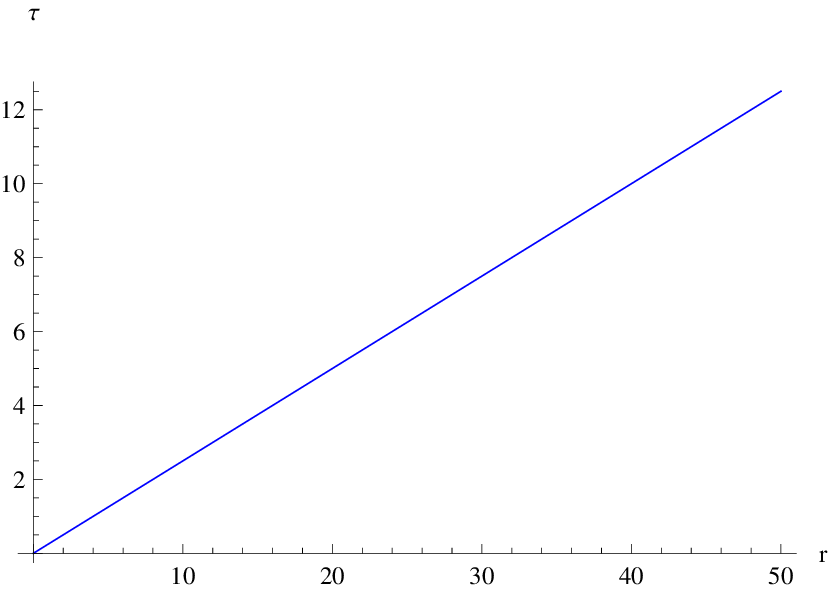,
width=0.5\linewidth} \caption{The left graph shows the behavior of
$r$ and $t$ from Eq.(\ref{10}) (when $m$=2 and $l$=1). The right
graph shows the relation of Eq.(\ref{12}) with $E=4$.}
\end{figure}
Now, from Eq.(\ref{6}), when $L=0$, the relation between $\tau$ and
$r$ is given by

\begin{equation}\label{11}
\left(\frac{dr}{d\tau}\right)^{2}=E^{2},
\end{equation}
the above equation implies that
\begin{equation}\label{12}
\pm E\tau=r.
\end{equation}
The changing of proper time $(\tau)$ and $r$ is shown in right graph
of figure \textbf{2}.
\subsection {Massive particle motion ($L$=1)}
Here we deal with the motion of massive particles when the
trajectories of the particles are radial direction of BH.
 From Eq.(\ref{6}), we get
\begin{equation}\label{13}
\left(\frac{dr}{dt}\right)^{2}=\left[f^{2}(r)-\frac{f^{3}(r)}{E^2}\right].
\end{equation}
Integration of this equation yields
\begin{eqnarray}\label{15}
 \pm t &=&
\frac{1}{18}\left[\frac{(42)^{2/3}m^{5/3}\log{(2l^{2/3}\sqrt[3]{m}+2^{2/3}r)}}{E^{2}
 l^{2/3}}-\frac{(42)^{2/3}\sqrt{3}m^{5/3}\tan^{-1}\left({\frac{2^{2/3}r}{l^{2/3}\sqrt[3]{m}}}-1\right)}{E^{2}l^{2/3}}\right.\nonumber\\
 &+&\left.\frac{12m^{2}r^{2}}{E^{2}(2l^{2}m+r^{3})}-\frac{(22)^{2/3}m^{5/3}\log{(2l^{4/3}m^{2/3}-2^{2/3}l^{2/3}\sqrt[3]{m}r+\sqrt[3]{2}r^{2})}}{E^{2}l^{2/3}}+9r(E^{2}+2)\right.\nonumber\\
 &+&\left.\frac{6(E^{4}+2E^{2}-1)m\log{(2l^{2}+r^{3})}}{E^{2}}\right].
\end{eqnarray}
The relation between $t$ and $r$ for the massive particles is shown
in the left graph of figure \textbf{3}.

From Eq. (\ref{4}), we get
\begin{equation}\label{16}
\left(\frac{dr}{d\tau}\right)^{2}=E^{2}-f(r).
\end{equation}
This implies that
\begin{equation}\label{17}
\pm \tau=
\int\frac{dr}{\sqrt{E^{2}-1+\frac{2mr^{2}}{r^{3}+2ml^{2}}}}.
\end{equation}
After integrating, we get
\begin{equation}\label{18}
\pm \tau= \frac{r}{2}+E^{2}r-\frac{1}{3}\log{(2l^{2}m+r^{3})}.
\end{equation}
This is the relation between proper time $(\tau)$ and $(r)$, which
is shown in right graph of figure \textbf{3}.
\begin{figure}
\epsfig{file=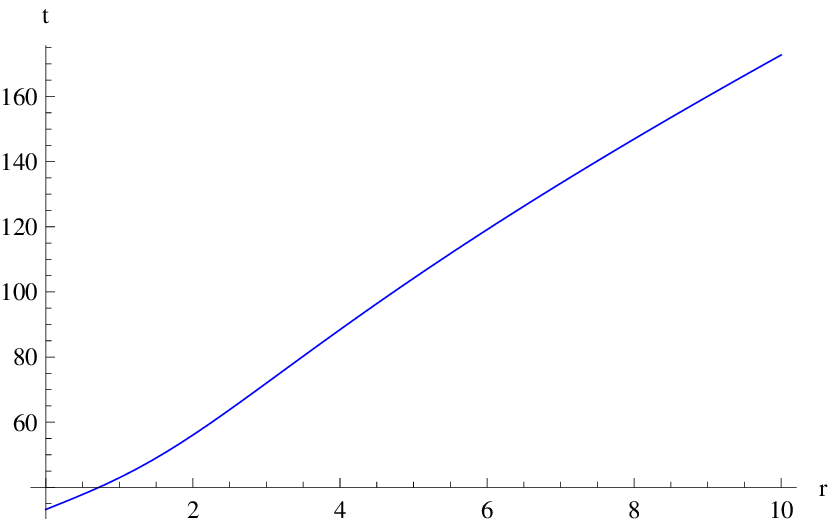, width=0.5\linewidth}\epsfig{file=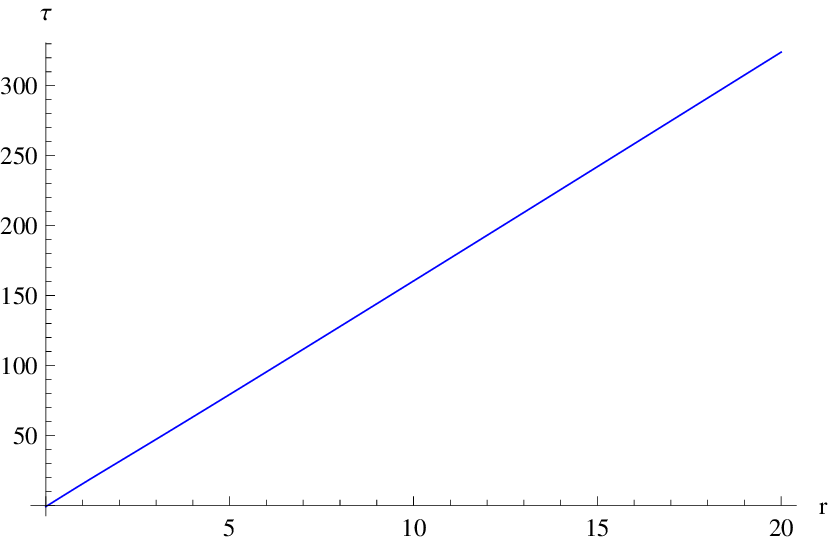,
width=0.5\linewidth}\caption{The left graph show the behave of $r$
and $t$ from Eq.(\ref{15}) (when $m$=2, $E=4$ and $l$=1). The right
graph show the relation of Eq. (\ref{18}) with ($m=2$, $l=1$ and
$E=4$)}
\end{figure}

\begin{figure}
\begin{center}
\epsfig{file=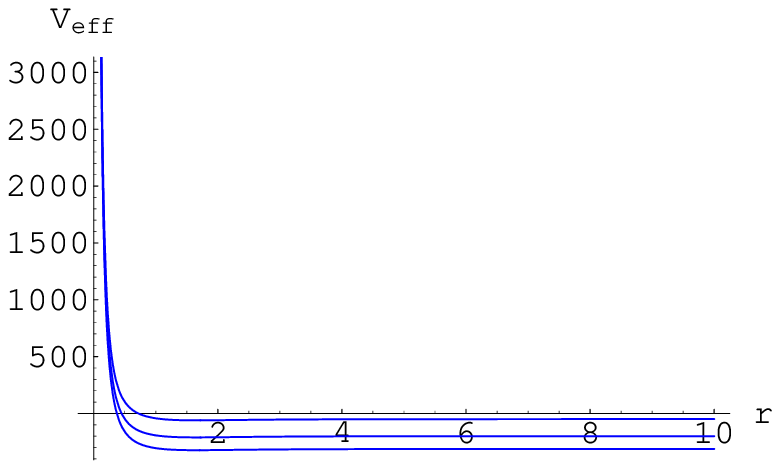, width=0.9\linewidth} \caption{This is shape of
effective potential $(V_{eff})$ for circular motion in Hayward
geometry. The graph has been plotted for $m=3$, $E=100, 200, 300,
J=10$ and $l=1$.} \epsfig{file=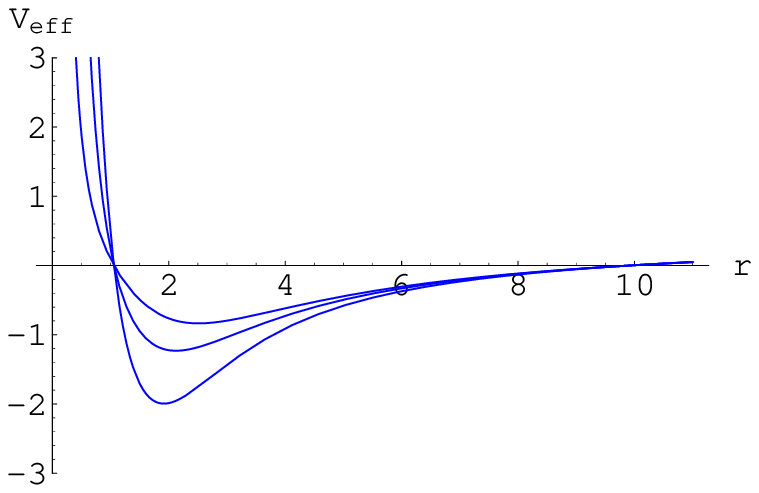, width=0.9\linewidth}
\caption{This is shape of effective potential $(V_{eff})$ for
circular motion in Hayward geometry. The graph has been plotted for
$m=3, E=0, J=10, 20, 30$ and $l=1.$ }
\end{center}
\end{figure}
\begin{figure}
\begin{center}
\epsfig{file=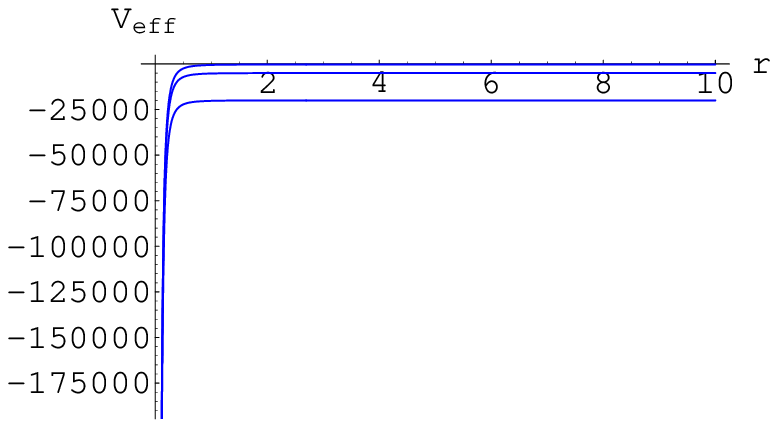, width=0.9\linewidth} \caption{This is the shape
of effective potential $(V_{eff})$ for circular motion in
Schwarzschild geometry ($l=0$ in Eq.(\ref{20})). The graph has been
plotted for $m=3$, $E=100, 200, 300$ and $J=10$.}
\epsfig{file=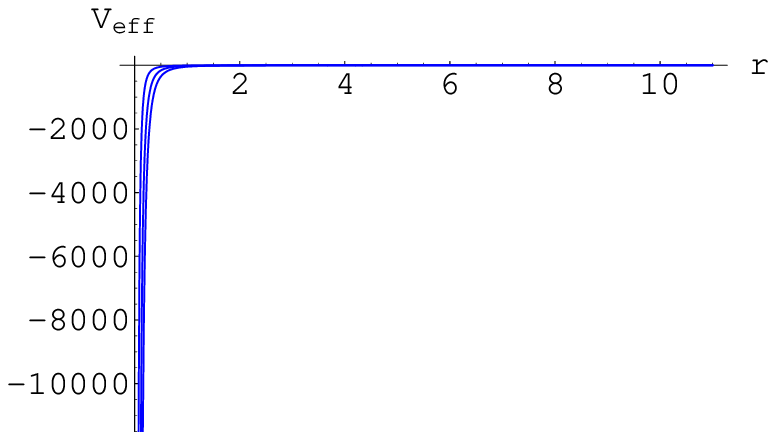, width=0.9\linewidth} \caption{This is the shape
of effective potential $(V_{eff})$ for circular motion in
Schwarzschild geometry ($l=0$ in Eq.(\ref{20})). The graph has been
plotted for $m=3$, $E=0$ and $J=10, 20, 30$.}
\end{center}
\end{figure}

\section{Effective Potential}

From the geodesic Eq.(\ref{3})
\begin{equation}\label{19}
{\dot{r}}^2\equiv\left(\frac{dr}{d\tau}\right)^{2}=E^{2}-f(r)\left(\frac{J^{2}}{r^{2}}+L\right).
\end{equation}
Comparing above equation with equation of motion
$\frac{\dot{r}^{2}}{2}+V_{eff}=0$, we get
\begin{equation}\label{20}
V_{eff}=-\frac{1}{2}\left[E^{2}-f(r)\left(\frac{J^{2}}{r^{2}}+L\right)\right].
\end{equation}
This leads to Schwarzschild BH effective potential when $l=0$ in
lapse function $f(r)$ of the Hayward BH.
\subsection {For Photon-like Particle ($L$=0)}
Consider the radial geodesic when $J=0$. Then, $V_{eff}$ is given by
\begin{equation}\label{21}
V_{eff}=-\frac{1}{2}E^{2}.
\end{equation}
This show that particle will behave like a free particle i.e., its
$V_{eff}=0$, for $E=0$ .

For circular geodesic case, $J\neq0$, the effective potential can be
written as,
\begin{equation}\label{22}
V_{eff}=-\frac{E^{2}}{2}+\frac{J^{2}}{2r^{2}}\left(1-\frac{2mr^{2}}{r^{3}+2l^{2}m}\right).
\end{equation}
In the limit $r\rightarrow0$, $V_{eff}$ attains a large value and
when $r\rightarrow\infty$, ${V_eff}\rightarrow-\frac{E^{2}}{2}$, and
graph shown in figures \textbf{4} and  \textbf{5}. By definition
horizon would occur at such values of radial position $r$ where
$f(r)=0$, so from Eq.(\ref{20}), we have
$V_{eff}=-\frac{1}{2}E^{2}<0$, for every $E$. In this case particles
have real velocity between the horizons.

\subsection {For Massive Particle ($L$=1)}
In this case effective potential is
\begin{equation}\label{23}
V_{eff}=-\frac{1}{2}\left[E^{2}-f(r)\left(\frac{J^{2}}{r^{2}}+1\right)\right].
\end{equation}
For $J=0$, we get same value of $V_{eff}$ as in case of photon-like
particle. It is interesting to study the motion of massive particle
for $E\geq0$. When $E=0$ and $J=0$, we get

\begin{equation}\label{23}
V_{eff}=\frac{1}{2}\left[1-\frac{2mr^2}{r^3+2l^2m}\right].
\end{equation}
This equation is nothing but the effective potential in term of
lapse function. The roots of $V_{eff}$ will be same as the roots of
$f(r)$ in figure\textbf{1}. Depending upon the values of the
parameters particles can move inside the BH. For $E\neq0$ and $J=0$,
we have $V_{eff}=-\frac{E^2}{2}+f(r)$. This has same interpretation
as in case of motion of photon at the horizon, i.e., $f(r)=0$ at
horizon and $V_{eff}<0$, particles have real velocity between the
horizons. In the limit $r\rightarrow\infty$, this potential attains
a constant value $V_{eff}\rightarrow\frac{1-E^2}{2}$.

Now, we consider the non-vanishing angular momentum case $J\neq0$,
with $E=0$, the effective potential becomes
\begin{equation}\label{23}
V_{eff}=\frac{1}{2}\left(1-\frac{2mr^2}{r^3+2l^2m}\right)\left(\frac{J^2}{r^2}+1\right).
\end{equation}
In this case the shape of potential coincides with the roots the
lapse functions which implies that particles tracing out the timlike
trajectories are captured by the BH, which makes these particles to
move in the circular orbits of fixed radius. The existence of minima
in figures \textbf{4} and  \textbf{5} indicate the stability of
circular orbit.

\section{Conclusion}
In this paper, we have  investigated the structure of timelike and
null geodesics of Hayward regular BH. The radial and non-radial
geodesics motion for both massless (photon) and massive particles
have been analyzed in explicitly. In case of radial motion, we have
determined the analytic solution of equation of motion, which
exhibit the relation between radial distance and time/proper time.
During the radial motion, the photon as well as massive particles
undergoes a small deviation in distance-time relation, while
distance-proper time relationships are linear as shown in figure
\textbf{2} and \textbf{3}. These relations are independent of
parameters of he BH and only depends on nature of geodesics (radial
or non-radial). The effective potential for radial motion of
photon-like particles in figures \textbf{4} and \textbf{5} imply
that these can behave as free particle if its energy is zero.

For the non-radial motion $J\neq0$, we have plotted the effective
potential for particular choice of parameters. When $J\neq0$, the
massless particles have real velocity and these are bounded to move
inside the horizons of the BH. For massive particle moving along
circular geodesic, when $E=0$, $J\neq0$, the effective potential has
same graph as the metric function plotted for the horizon in figure
\textbf{1}. Hence particle moving along timelike curved path are
attracted by the gravity of BH in such a way that particles
continues to circulate around BH in particular orbits. The minima in
the effective potential of non-radial motion of massive particles in
implies that the circular orbits are stable.

The general relativity was experimentally tested in a weak
gravitational field. Testing general relativity in strong
gravitational field requires the investigation of some astrophysical
phenomena near compact objects, such as BH or a neutron star.
Astrophysical observations of galaxies imply that their centres are
occupied by the massive dark objects. The arguments suggest that
these are supermassive black holes. The observational targets to
test the Einstein theory of relativity in a strong gravitational
field is gravitational lensing. The basic tool for studying the
gravitational lensing near a massive object is the geodesic study of
that object. The gravitational lensing through Schwarzschild BH in
the weak gravity region ( i.e.; for small deflection angle) is
well-known. Kling et al. (2000) have adopted the numerical approach
for gravitational lensing theory based on the approximate solutions
of the geodesics equations. Virbhadra and Ellis (2000) have
introduced a lens equation that allows for the large bending of
light near a black hole, they model the Galactic supermassive
Schwarzschild lens and study point source lensing in the strong
gravitational field region. The geodesic study presented in this
paper will be helpful for studying the gravitational lensing of
Hayward BH. This will be done explicitly in an other investigation.

\vspace{0.25cm}

{\bf Acknowledgment}

\vspace{0.25cm}
 We highly appreciate the fruitful comments of the anonymous referee
 for the improvements of the paper.

\end{document}